\documentclass[12pt]{article}
\usepackage{moreverb}

\usepackage{amsmath}
\usepackage{graphicx}
\usepackage{enumerate}
\usepackage{amssymb,mathtools,xcolor} %
\usepackage{hyperref}
\usepackage{verbatim}
\usepackage{fancyvrb} % indent code in Verbatim
\usepackage[numbers]{natbib}
\usepackage{authblk} % customize author and affiliation
\usepackage{multirow}
\usepackage{booktabs}
\usepackage{upgreek}  % upright greek letter
\usepackage{setspace} 

\usepackage[inner=0.8in,outer=0.8in,bottom=1in, top=1in]{geometry}

\newtheorem{remark}{Remark}

\title{\bf Meta-analysis of Censored Adverse Events}
\author[1]{Xinyue Qi}
\author[2,*]{Shouhao Zhou}
\author[1]{Christine B. Peterson}
\author[3]{Yucai Wang}
\author[2]{Xinying Fang}
\author[4]{Michael L. Wang}
\author[2,5]{Chan Shen}
\affil[1]{\footnotesize Department of Biostatistics, The University of Texas MD Anderson Cancer Center, Houston, TX}
\affil[2]{Department of Public Health Sciences, Penn State College of Medicine, Hershey, PA}
\affil[3]{Division of Hematology, Mayo Clinic, Rochester, MN}
\affil[4]{Department of Lymphoma, The University of Texas MD Anderson Cancer Center, Houston, TX}
\affil[5]{Department of Surgery, Penn State College of Medicine, Hershey, PA}

\affil[*]{Corresponding author: shouhao.zhou@psu.edu}
\date{}

\begin{document}
\maketitle
\begin{abstract}
Meta-analysis is a powerful tool for assessing drug safety by combining treatment-related toxicological findings across multiple studies, as clinical trials are typically underpowered for detecting adverse drug effects. However, incomplete reporting of adverse events (AEs) in published clinical studies is a frequent issue, especially if the observed number of AEs is below a pre-specified study-dependent threshold. Ignoring the censored AE information, often found in lower frequency, can significantly bias the estimated incidence rate of AEs. Despite its importance, this common meta-analysis problem has received little statistical or analytic attention in the literature. To address this challenge, we propose a Bayesian approach to accommodating the censored and possibly rare AEs for meta-analysis of safety data. Through simulation studies, we demonstrate that the proposed method can improves accuracy in point and interval estimation of incidence probabilities, particularly in the presence of censored data. Overall, the proposed method provides a practical solution that can facilitate better-informed decisions regarding drug safety.
\end{abstract}


\section{Introduction} % rare event
\label{sec1}
This paper focuses on statistical modeling strategies for analyzing drug safety data across multiple studies. Our motivation stems from a systematic review of treatment-related adverse events (AEs) of programmed cell death 1 (PD-1) and PD-1 ligand 1 (PD-L1) inhibitors for cancer immunotherapy \citep{wang2019treatment}. The PD-1 pathway is up-regulated in many tumors and in their microenvironment. Blockade of this pathway with antibodies to PD-1 or its ligands has led to remarkable clinical responses in various types of cancer \citep{le2015pd}, and is considered one of the most important breakthroughs in the treatment of cancer. These novel immune checkpoint inhibitors are clinically less toxic than traditional cancer treatments such as chemotherapy and radiation therapy, but can occasionally cause serious and sometimes life-threatening immune-related AEs. Because individual clinical trials are often underpowered for detecting adverse drug effect, combining evidence from multiple studies is an attractive approach to examine the toxicological profile of PD-1 and PD-L1 inhibitors.

Meta-analysis synthesizes findings from multiple independent clinical studies and provides a more powerful analysis than from a single study \citep{sutton2002meta,higgins2019cochrane}.
However, one unique challenge in meta-analysis of drug safety is the incompleteness of AE reporting, which hinders the accuracy to quantify and understand the incidence of treatment-related AEs \citep{us2015adverse}. A significant portion of treatment-related AEs, despite having been counted in individual clinical trials, may not be publicly reported due to the rarity of such events.  In the motivating anti-PD-1/PD-L1 example, many AEs were missing if their study-level observed frequencies were lower than a pre-determined reporting cutoff (e.g. 3\% or 5\% of the study sample size). {\color{black} Table \ref{tab:1} provides several sample studies on reporting of pneumonitis, a type of AE regarding lung inflammation, to illustrate the AE data structure in a typical meta-analysis. For instance, Powles \textit{et al.} \citep{powles2018atezolizumab} conducted an open-label randomized controlled trial to assess the the safety and efficacy of Atezolizumab. Among 459 patients treated with Atezolizumab, the AE reporting criteria was $\ge$ 5\% for all-grade and $\ge$ 2\% for grade 3 or higher, suggesting the missing pneumonitis count could be any number from 0 to 22 for all-grade and from 0 to 9 for grade 3-5.} In another small multicenter study of Pembrolizumab, Nanda \textit{et al.} \citep{nanda2016pembrolizumab} only listed the exact count of all grade AEs occurring in at least two patients (implying the reporting cutoff of 1 for censored AEs). Because pneumonitis wasn't reported in their results, it implies 0 or 1 case of all grade pneumonitis rather than the absence of pneumonitis. Likewise, Nanda \textit{et al.} \citep{nanda2016pembrolizumab} only listed grade 3-5 AEs occurring in at least one patient, which indicates a reporting cutoff of 0 and implies that the actual grade 3-5 pneumonitis count, though unreported, was 0. Subsequently, if the analysis was conducted only based on the likelihood of the reported AE frequencies but ignoring all the left censored data, the inferences on incidence rates could be significantly overestimated.

\begin{table}[!ht]
        \centering
        \scalebox{0.75}{\begin{tabular}{ccccccccc}
        \hline
        \hline
        &  & No. of Treated & & Cutoff & Pneumonitis & Cutoff & Pneumonitis \\
        &  Study Source &  Patients &  Drug &  (all grade) & (all grade) & (grade 3-5) & (grade 3-5) \\
        \hline
        & 2015-Robert-N Engl J Med2 \citep{robert2015pembrolizumab} & 206  & Nivolumab &  4 &  -  & 4 & - \\
        & 2018-Lee-Lung Cancer \citep{lee2018nivolumab} & 100 &  Nivolumab &  0 &  3  & 0 & 2  \\ 
        & $$\vdots$$ & & $$\vdots$$ &  &  &  &  \\
        & 2016-Nanda-J Clin Oncol \citep{nanda2016pembrolizumab} & 32 & Pembrolizumab &  1 &  - & 0 & - \\
        & 2017-Hsu-J Clin Oncol \citep{hsu2017safety} & 27 & Pembrolizumab &  1 &  3 & 1 & 2 \\ 
        & $$\vdots$$ & & $$\vdots$$ &  &  &  &  \\
        & 2018-Horn-Eur J Cancer \citep{horn2018safety} & 89 & Atezolizumab &  4 & - & 4 & -  \\ 
        & 2018-Powles-Lancet \citep{powles2018atezolizumab} & 459 & Atezolizumab &  22 &  - & 9 & -  \\ 
        & $$\vdots$$ & & $$\vdots$$ &  &  &  &  \\
        \hline
        %& $i$ & $N_i$ &  $j$ & $Z_{ij,1}$ & $Z_{ij,2}$ & $X_{ij,1}$ & $X_{ij,2}$ & $X_{ij,3}$ &  $c_{i}$ & $Y_{ij}$\\ 
        \end{tabular}}
        \caption[]{A sample of AE observations for the motivating real-data application in Section \ref{sec4}. ``-'' indicates the missing data (i.e., unreported in publicly available data). The left-censored cutoffs for all grade AEs and grade 3-5 AEs could be different within a study. Given the cutoff of 0 for grade 3-5 AEs in \textit{2016-Nanda-J Clin Onco} study, the actual grade 3-5 pneumonitis count, though unreported, was exactly 0. In the proposed Bayesian modeling framework, identical inferences are generated by either treating it as a censored observation or observed.}       
        \label{tab:1}
\end{table}

Although censored AE data is a common issue in drug safety analysis, it has been largely overlooked in the development of meta-analysis methodology. In fact, the focus of attention in meta-analyses of treatment-related AEs has been mostly on the rarity of such events \citep{bhaumik2012meta}. {Standard methods to model binary patient outcomes such as AE incidences rely on either approximation methods based on the normal distribution or exact methods based on the binomial distribution \citep{hamza2008binomial}. When the observed events are rare, approximation approaches may provide poor estimates of the true incidences and lead to significantly biased results \citep{luft1993calculating, carriere2001good,  bradburn2007much, lane2013meta}.}
Some recent efforts have been made to overcome this limitation, including the Poisson random-effect model to estimate relative risk between two treatment groups \citep{cai2010meta}, and asymptotically unbiased estimation for the treatment effect and heterogeneity parameter in the random-effect model \citep{bhaumik2012meta}. However, these methods were all developed for meta-analysis without missing data.

Most of the research on missing data in meta-analysis focuses on situations when the estimate from the whole study is missing \citep{pigott1994methods}, or on the analysis of treatment efficacy in different patterns \citep{white2008allowingA, white2008allowingB, higgins2008imputation, mavridis2014addressing}. Due to the lack of appropriate analytic methods to address the problem of censored AE data , in current meta-analytic applications, most studies either totally ignored the AEs with low incidence, or completely discarded the studies with missing AE data \citep{silva2006statin}, contributing to substantial publication selection bias or estimation error. 

In this paper, we propose a general Bayesian approach to model censored AE data in a meta-analysis, with an aim to deliver exact inference in the estimation of AE incidence with proper uncertainty quantification. The rest of the article is organized as follows. In Section \ref{sec2}, we present the Bayesian modeling framework and implement the proposed approach in Just Another Gibbs Sampler (JAGS) with a tailored presentation for model assessment.  In Section \ref{sec3}, we conduct numerical studies under different censoring scenarios by comparing the proposed Bayesian model of censored data with other popular methods. Real data meta-analysis results demonstrating the advantage of the proposed approach are presented in Section \ref{sec4}. Lastly, some concluding remarks and discussions are provided in Section \ref{sec5}.

\section{Methods}
\label{sec2}

We propose a Bayesian \underline{m}eta-analysis of \underline{a}dverse dru\underline{g} \underline{e}ffects with \underline{c}ensored data (MAGEC) framework, to accommodate censored AE data in meta-analysis. It incorporates cumulative probabilities of the partial information contained in the censored data into the likelihood function, such that the proposed approach can yield proper parameter estimation and statistical inference. 

\subsection{Modeling of Censored Adverse Events}
Let $Y_j$, $j=1,\cdots,J$, denote {a safety} response of interest in the $j$th study, which are collected in individual trials but may not be fully reported. For censored outcomes, the censoring mechanism can be defined by bounding variables $(A,B)$, with semi-closed boundaries $(A_j,B_j]$ for response variable $Y_j$. Here, both bounding variables could be covariate-dependent. Often, we assume that the outcome model and the censoring mechanism are independent, so the censoring mechanism does not contribute to the inference and model estimation. It is a fundamental assumption for censored data behind most statistical methodologies {\citep{zhang2010interval}}. In survival analysis, it is also known as noninformative censoring.  Assume the random variable $Y$ has the right continuous cumulative distribution function $F_{Y}(y)=P[Y\leq y]$.  Denote $f_{Y}(y)$ the probability density/mass function of $F_{Y}(y)$. 

In a meta-analysis of safety data, the frequency of adverse events may not always be reported. For example, left censoring occurs when some severe (grade 3 or higher) events are not observed due to low incidence. In this case, the cutoff boundaries are not random but fixed and study-specific, which spontaneously satisfies the noninformative censoring assumption.
Denote the fixed cutoff by $c_j$ for each study. The number of subjects having a specific event in the $j$th study $Y_j$ follows a binomial distribution with study-level sample size $n_j$ and AE incidence probability $\theta_j$
$$
    Y_j \sim Bin(n_j, \theta_j).
$$
Therefore, the likelihood function for both fully observed and censored events can be written by:
\begin{equation} % eq#
	\begin{split}
	\mathcal{L} %&= \prod_{o=1}^{O}f_{Y}(y_o) \prod_{l=1}^{L}[F_{Y}(b_l)-F_{Y}(a_l)] \prod_{r=1}^{R}[F_{Y}(b_r)-F_{Y}(a_r)] \\
	&= \prod_{o=1}^{O}f_{Y}(y_o) \prod_{l=1}^{L}[F_{Y}(c_l)-0] \prod_{r=1}^{R}[F_{Y}(n_r)-F_{Y}(c_r)] \\
	&= \prod_{o=1}^{O}f_{Y}(y_o) \prod_{l=1}^{L}F_{Y}(c_l)\prod_{r=1}^{R}[1-F_{Y}(c_r)] \\
	&= \prod_{o=1}^{O}f_{Y}(y_o) \prod_{l=1}^{L}\sum_{k_l=0}^{c_l}f_{Y}(k_l) \prod_{r=1}^{R}[1-\sum_{k_r=0}^{c_r}f_{Y}(k_r)],  
	\label{eq2}
	\end{split}
\end{equation}
where $O$ is {the number} of fully-observed AE outcomes, $L$ {the number} of left-censored outcomes, and $R$ {the number} of right-censored outcomes. $c_l$ and $c_r$ are cutoff values for left-censored and right-censored data, respectively. 

The likelihood (\ref{eq2}) can simultaneously handle both cases of left- and right-censoring for meta-analysis of safety data. In meta-analysis of the grade 3-5 AEs, we often only observed left censoring due to reporting cutoffs.
In meta-analysis of the all-grade AEs, right censoring may also occur when some studies report grade 2 or higher AEs only rather than all-grade AEs \citep{brahmer2010phase}.
If $Y_j$ is left-censored, then $Y_j$ lies in the semi-closed interval ($A_l=0^-$, $B_l=c_l$], where $c_l$ is the corresponding cutoff value for left-censored data. If ${Y_j}$ is right-censored, then $Y_j$ lies in the semi-closed interval ($A_r=c_r, B_r=n_r$], where $c_r$ is the corresponding cutoff value for right-censored data and $n_r$ is the total number of subjects in the $r$th study. This is specified to be consistent with the JAGS model implementation in Section \ref{algorithm}. Compared with multiple imputation approaches that often requires the assumption of normality and involves intensive computation for Bayesian modeling \citep{kalaycioglu2016comparison}, the proposed approaches are computationally efficient to model incomplete data. 

The incidence probability can be decomposed through the link function $g(\theta)$,
\begin{equation}
	g(\theta_j) = \textsf{logit}(\theta_j) = \mu + \alpha_j + \boldsymbol{X_j\beta},
	\label{eq3}
\end{equation}
where $\boldsymbol{X_j}$ is a design matrix for study-level covariates from the $j^{\text{th}}$ study, and we consider the $\textsf{logit}$ link as the default link function. Given the overall mean $\mu$ and the study covariate effects $\boldsymbol{\beta}$ associated with $\boldsymbol{X_j}$, we assume the study-specific random effect $\alpha_j$ is conditionally independent with mean $0$ and variance $\sigma_\alpha^2$. 

For nonhierarchical intercept $\mu$ or slope parameters in $\boldsymbol{\beta}$ in the linear term of logistic model, a symmetric Cauchy distribution with center 0 and scale 2.5 was suggested \citep{gelman2008weakly}. For random effects in the linear term, which include the effects of various study-level factors and study-specific effects $\alpha_j$, we assume they follow mean-zero normal distribution with different variances, with the prior distributions of standard deviation parameters following half-Cauchy distribution and a finite scale parameter of 10 or 25 \citep{gelman2006prior}. In practice, the normality assumption can also be replaced by using heavy-tailed distributions such as $t$-distribution. In the illustrative anti-PD-1/PD-L1 example, the meta-analysis included 125 studies with a total of 20,218 patients. To identify possible sources of heterogeneity between studies, the following pieces of study-level information were also extracted: trial name, number of treated patients, selected immunotherapy drug, dosing schedule, cancer type, AE frequency (which could be missing), and the study-specified censoring cutoff for AE reporting. In particular, the study-level factors could include the drug type or drug/dose combinations, and/or cancer type, such that it is possible to estimate the AE incidence of a new drug or a new cancer type in a future trial with Bayesian credible interval.

Lastly, the model specification of the proposed Bayesian modeling framework only involves discrete data in the likelihood function, while the prior distribution of parameters is proper. As a result, it is easy to verify the property of Bayesian posterior inference. Specifically, we can have the following remark:
\begin{remark}\label{RM:1}
The posterior distribution of Bayesian models in the MAGEC framework is always proper. 
\end{remark}
Overall, the proposed Bayesian framework can help to streamline the modeling process of drug safety data and enhance the reporting reliability of meta-analysis.

\subsection{Model Implementation using JAGS} 
\label{algorithm}

To carry out posterior inference and assess the performance of the proposed Bayesian model, we apply Just Another Gibbs Sampling (JAGS) to generate samples from the posterior distribution. JAGS makes Bayesian hierarchical models easy to implement using Markov Chain Monto Carlo (MCMC) simulation \citep{plummer2003jags} in \textsf{R} and other computational software. In the presence of censored data in the response variable, an existing function, known as \textsf{dinterval} distribution, is commonly used to model censored data \citep{kruschke2014doing, plummer2017jags}. However, such model specification for censored data in JAGS intrinsically yields a mis-specified deviance function \citep{sourceforge2012}. {More technical details are provided in Qi \textit{et al.} (2022) \cite{qi2022bayesian}.}

Alternatively, we apply a simple but effective approach to censored data specification. To facilitate model implementation for censored observations (when $\delta_j = 0$) and avoid the miscalculation of deviance via the $\texttt{dinterval()}$ function in JAGS. Here, we utilize the idea of data augmentation by introducing ancillary indicator variables $W_{j}$. Each $W_{j}$ separates the left-censored from right-censored observations ($W_{j} =1$ if left-censored, $0$ if right-censored). By assuming $W_{j}$ follows a Bernoulli distribution, we have the density function  
\begin{align}
f_W(w_j) & = F_{Y}(c_l)^{w_j}[1-F_{Y}(c_r)]^{1-w_j} \nonumber \\
   & = \left [ \sum_{k_l=0}^{c_l}f_{Y}(k_l)\right ]^{w_j} \left[1-\sum_{k_r=0}^{c_r}f_{Y}(k_r)\right]^{1-w_j},
\end{align}
which exactly matches the second and third terms for censored observations in (\ref{eq2}), with the cumulative binomial distribution of incidence probability of AE in the $j$th study, restricted by a pre-determined study-level cutoff value $c_{j}$. This approach can also be extended to interval-censored data \citep{qi2022bayesian}.

A JAGS model specification for the application is provided in \textit{Appendix} for illustrative purposes. Together with fully observed data that follow a binomial distribution, the full likelihood implemented in a JAGS model is, in fact, identical to the exact likelihood of observed and censored cases in (\ref{eq2}). 
This creates the correct focus of model parameters and produces the proper posterior samples, as well as simultaneously computes the correct deviance for model selection. For example, by calling the deviance module in JAGS, correct deviance information criterion (DIC) \citep{spiegelhalter2002bayesian}, penalized deviance \citep{plummer2008penalized}, posterior averging information criterion (PAIC) \citep{zhou2023posterior}, or widely applicable information criterion (WAIC) \citep{watanabe2010asymptotic, vehtari2017practical} can be conveniently derived to assess candidate models for Bayesian model selection. It is important and beneficial to select the best model, especially in the presence of complicated model features. Also, from the point of view of convenience, the sampling procedure can smoothly start with default initial values for the censored data. The standard diagnostic tools can be applied to assess the posterior convergences of the model \citep{gelman1995bayesian}.

\section{Simulation} % Section 3
\label{sec3}
In this section, we conduct a simulation study to assess the performance of the proposed Bayesian model for estimating the incidence rates in the meta-analysis of adverse events (AEs) with censored information. We compare it with that of other popular methods applied under a standard setting \citep{silva2006statin}. {Other than the proposed Bayesian model, MAGEC, all other methods do not account for censoring.}

\subsection{Settings}
The total number of studies for each probability level of AE incidence is fixed at 10 ($J = 10$) to reflect the typical number of studies in a meta-analysis for subgroups. We also consider $J =$ 25 and 50 as comparisons. The outcome of interest, the number of AEs for each study, is generated from a binomial distribution with the number of patients ($n = 100$) and probability of events ($p = 0.2, 0.05$, and $0.01$, respectively). The probability of incidence is designed to cover the typical range of the probability levels of AE incidence. {The data generation incorporates heterogeneity between studies into the incidence probability of AE. } {To be specific, the coefficient of study effect was generated from a normal distribution at mean 0 and standard deviation of 0.2.}

To assess the performance of the proposed model that incorporates both observed and censored data, we consider {{six}} scenarios: (1) no censoring; (2) a low percentage (20\%) of {left-}censoring for {all three incidence levels}; (3) a {moderate} percentage (40\%) of {left-}censoring;  (4) a high percentage (60\%) of {left-}censoring;  {(5) a low/moderate percentage (20\%/40\%) of right-censoring; and (6) a moderate percentage (40\%) of left-censoring with a low percentage (20\%) of right censoring}.  In Scenario 1, the number of events for all studies is fully observed. In Scenarios 2, 3 and 4, which include observations {that could only be informatively left-censored for zero and low incidence counts} to mimic real-world cases. Therefore, in Scenario 2, we treat the 20\% of AE data with {the lowest observed} incidence as censored data and the 80\% of AE data that have a higher incidence as observed data. Similarly, in Scenario 3, in order to stress test the robustness of estimation in a more extreme case of censoring, 40\% of AE data with low incidence are treated as censored and the remaining 60\% are treated as observed. {Lastly, in Scenario 4, the top 40\% of studies are treated as observed data, and the remaining 60\% are treated as censored data.} {Accordingly, the study-specific left-censored cutoff value is selected from the sorted outcomes in a descending order based on the percentage of left-censoring after data generation. For example, in the scenario of 60\% left-censoring, the cutoff value corresponds to the $((1-60\%)J + 1)^{th}$ outcome among sorted values.} {However, the right-censoring in AE reporting is a different censoring mechanism (e.g., the reporting of grade 2 or higher AEs rather than all-grade AEs) and likely from the random selection of studies. Therefore, in Scenario 5, 20\% of studies are randomly treated as right-censored; in Scenario 6, 40\% of AE data with low incidence are treated as left-censored data first and then randomly selecting one third of the remaining 60\% (e.g., 20\% among all) of data as right-censored.} {The study-specific right-censored cutoff value is generated from a binomial distribution by assuming the probability of grade 2 or higher AE is 50\% of all-grade AE incidence.} 

We compare the proposed model, {MAGEC}, with four other methods: the pooled estimation method after continuity correction (PEM) \citep{sweeting2004add, jewell2003statistics}, the normal approximation method (NAM) \citep{peizer1968normal}, the logistic regression method (LRM), and the normal approximation method with robust variance estimator (RVE) \citep{ma2016meta}. In {MAGEC}, following the recommendation of the weakly-informative prior distribution for logistic regression models \citep{gelman2008weakly}, we assign a half-Cauchy prior distribution on the standard deviation of study effects. In PEM \citep{sweeting2004add}, we pool observations by {incidence level} and the 95\% nominal confidence intervals (CIs) for {AE incidence levels} are calculated by the exact binomial test. In NAM, a standard method in practice \citep{peizer1968normal}, we use a normal likelihood procedure to estimate the incidence rate by taking the inverse logit of the observed logit incidence \citep{hamza2008binomial} of each {incidence level} weighted by its within-level variance. In LRM, we estimate the {AE incidences} by an exact method through fitting a generalized linear model with a logit link. In addition, we compare the performance of NAM with and without robust variance estimators \citep{ma2016meta}. Therefore, in RVE, instead of Fisher information, we implement the sandwich estimator of variance into NAM to improve the robustness of the statistical inference on incidence rates.

We assess model performance in terms of accuracy in the estimation of the AE incidences.  More specifically, we compare the five methods using the following metrics:  mean absolute deviation (MAD), root mean squared error (RMSE) of the point estimates (for {MAGEC}, we use the posterior median), and coverage probability (CP) of the 95\% nominal confidence interval (for {MAGEC}, we use the 95\% posterior credible intervals), of the incidence rate parameters of interest in each scenario.

\subsection{Simulation Results}
\label{sec3.2}
\begin{figure}[!htb] 
	\centering
	\includegraphics[scale=0.43]{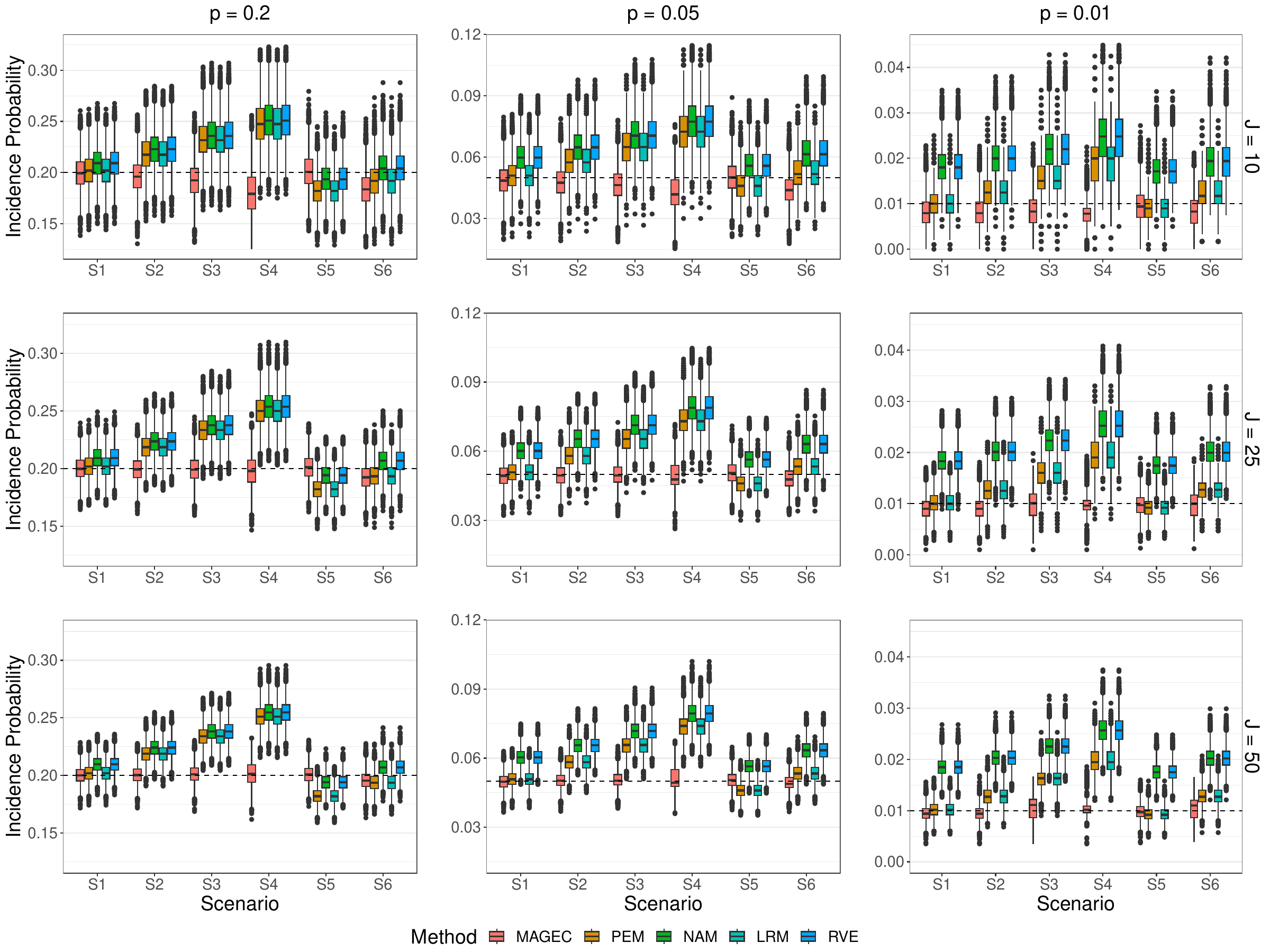} % 0.45
	\caption[Boxplot of point estimators of drug incidence rates for five methods under {four} scenarios.]{ Boxplot of point estimators for AE incidence using five methods. We compared the {proposed Bayesian model (MAGEC)},  pooled  estimation  method  after continuity correction (PEM), normal  approximate  method  (NAM),  logistic  regression  model  (LRM), and  normal approximate method with robust variance estimation (RVE) under {different total number of studies ($J=$ 10, 25, 50)} and {six} scenarios: (S1) 0\% {left} censoring; (S2) 20\% {left} censoring; (S3) 40\% {left} censoring; (S4) 60\% {left} censoring; {(S5) 20\% right censoring; and (S6) 40\% left censoring \& 20\% right censoring. }}
	\label{fig:1}
\end{figure}

\begin{figure}[!htb]
	\centering
	\includegraphics[scale=0.45]{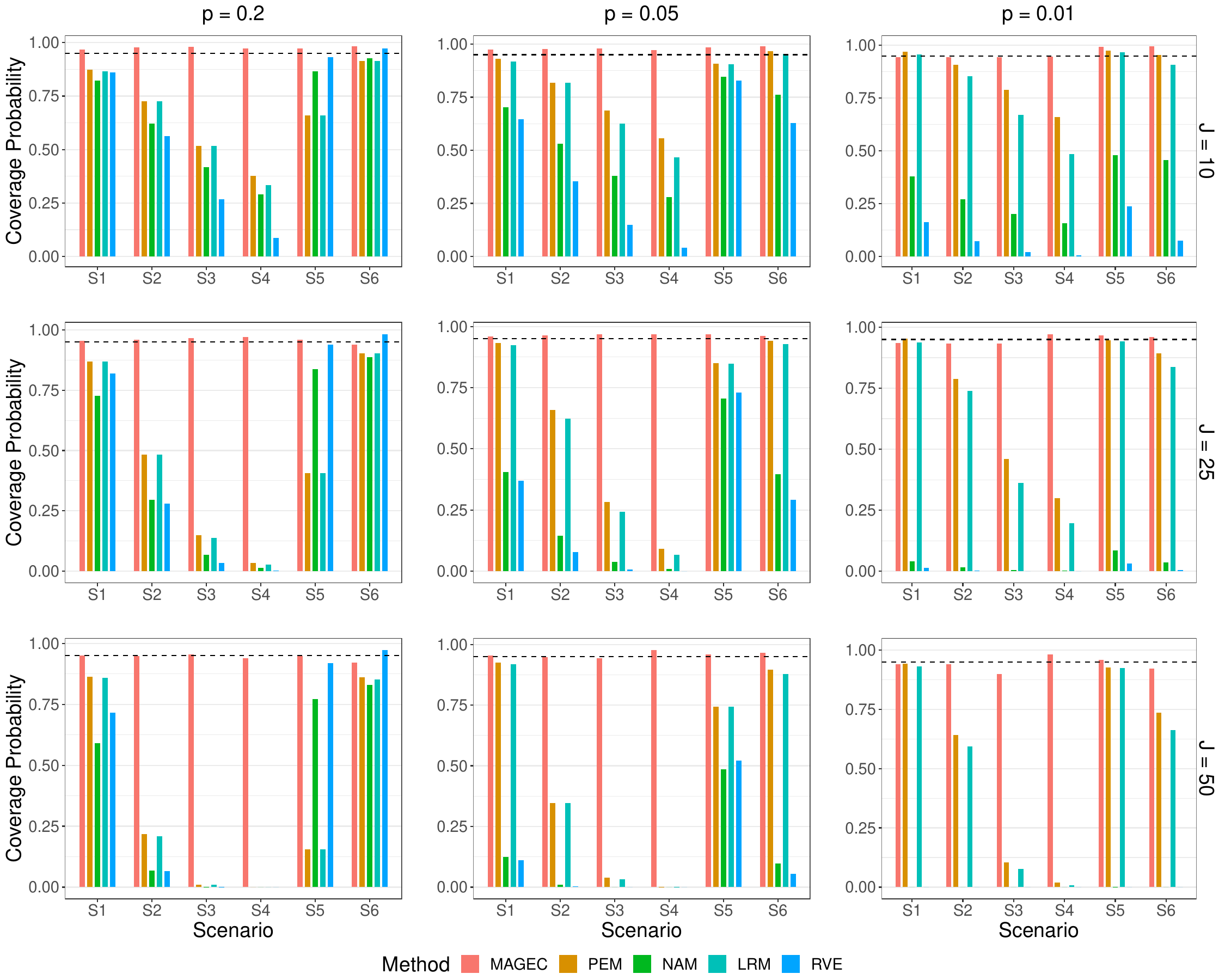} % 0.45
	\caption[Coverage probabilities of drug incidence rates for five methods under {four} scenarios.]{Coverage probabilities (CPs) of AE incidences using five methods. We compared the {proposed Bayesian model (MAGEC)},  pooled  estimation  method  after continuity correction (PEM), normal  approximate  method  (NAM),  logistic  regression  model  (LRM),  as  well  as  normal approximate method with robust variance estimation (RVE) under different total number of studies ($J=$ 10, 25, 50) and {six} scenarios: (S1) 0\% {left} censoring; (S2) 20\% {left} censoring; (S3) 40\% {left} censoring; (S4) 60\% {left} censoring; {(S5) 20\% right censoring; and (S6) 40\% left censoring \& 20\% right censoring.}}
	\label{fig:2}
\end{figure}

The results are based on 10,000 simulated data sets. For each method, we repeated the same data generation procedure in order to be able to compare results across methods. Figure \ref{fig:1} gives boxplots for the point estimates of incidence probabilities by scenario, method and the total number of studies. CPs of three parameters of interest by scenario, method, and the total number of studies are displayed in bar charts in Figure \ref{fig:2}. In Table \ref{tab:2}, performance in terms of both MAD and RMSE of incidence rates based on the five methods are shown for the {six} data censoring scenarios. %{Table \ref{app:b} that presents two additional data censoring scenarios is available in \textit{Appendix B}. }

\begin{table}[!htb]
	\centering \vspace{-20pt}
	\includegraphics[scale=0.67, trim={0cm 1.1cm 0cm .6cm},clip]{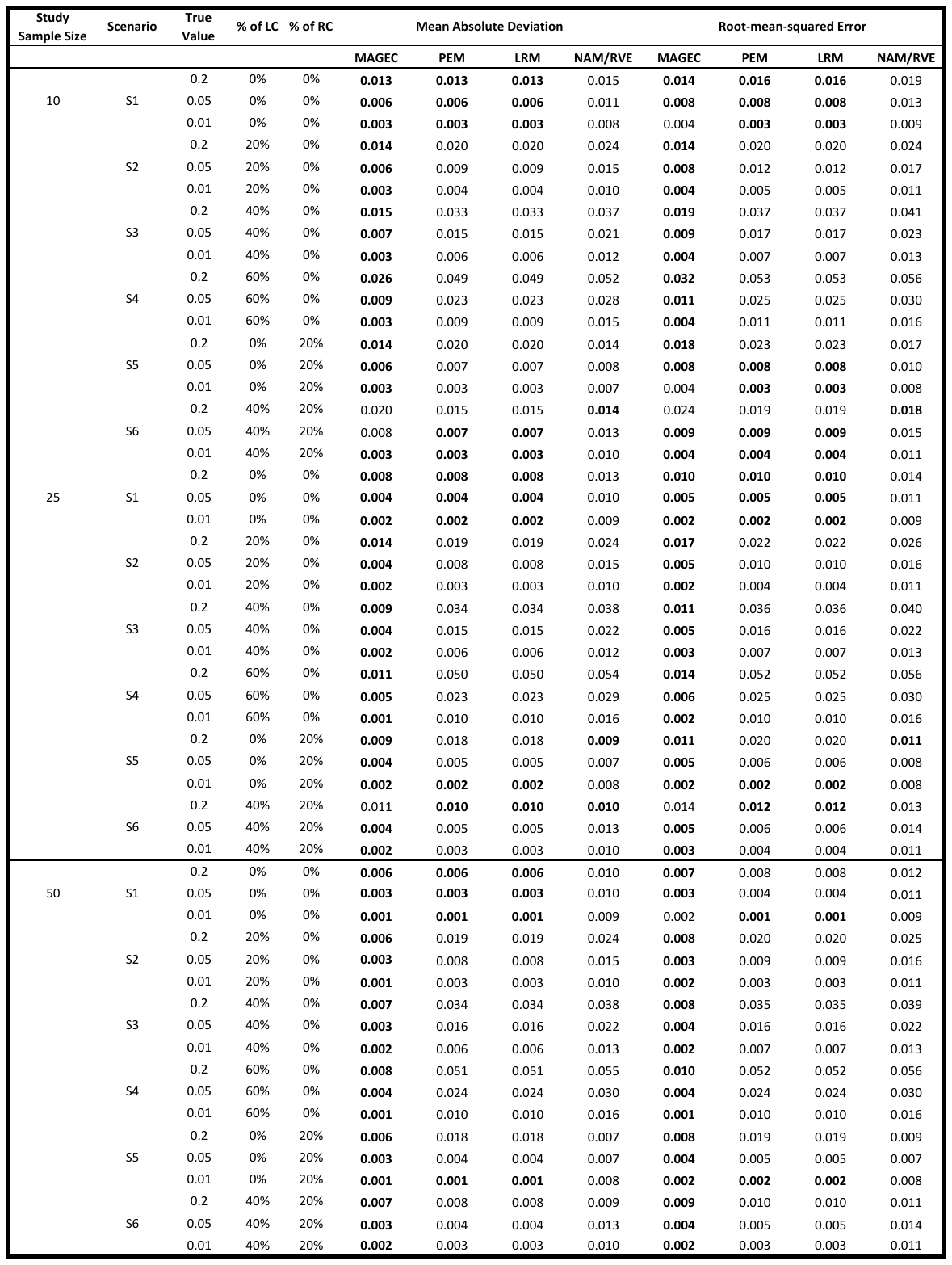} % 0.78
	\caption[Mean absolute deviations and root mean square errors of AE incidences for five methods under {four} scenarios.]{Mean absolute deviations (MADs) and root mean square errors (RMSEs) of drug incidence probabilities for five methods. We compared the {proposed Bayesian model (MAGEC)},  pooled  estimation  method  after continuity correction (PEM), normal  approximate  method  (NAM),  logistic  regression  model  (LRM),  as  well  as  normal approximate method with robust variance estimation (RVE) under different study sample sizes ($J=$ 10, 25, 50) and {six} scenarios: (S1) 0\% {left} censoring; (S2) 20\% {left} censoring; (S3) 40\% {left} censoring; (S4) 60\% {left} censoring; (S5) 20\% right censoring; and (S6) 40\% left censoring \& 20\% right censoring. {LC = left censoring; RC = right censoring. }}
	\label{tab:2}
\end{table}

When there is no censoring (Scenario 1), the proposed method ({MAGEC}) has CPs (Figure {\ref{fig:2}}), MADs, and RMSEs ({Table \ref{tab:2}}) for the incidence probabilities that are almost identical to those of PEM and LRM. Of the five methods compared, PEM can be considered the gold standard/benchmark for both interval and point estimation. Our results indicate that {MAGEC} is not inferior to PEM. They also indicate that the CP for each {probability level of AE incidence} obtained from NAM {and RVE} appeared to be unstable in estimating the incidence probabilities of less frequent events compared with the other methods. The point estimates of the incidence probabilities in both NAM and RVE are overestimated in Scenario 1. This finding is consistent with arguments mentioned in the normal approximation for rare events \citep{carriere2001good} and biased results of estimation for rare events using normal approximation \citep{luft1993calculating}. 

When 20\% of data are censored (Scenario 2), the proposed method ({MAGEC}) performs better than the others in estimating incidence probabilities; {with an increasing total number of studies}, its performance in Scenario 2 is comparable to that in Scenario 1. Because censored observations are ignored under the other four methods (PEM, NAM, LRM, and RVE), it is unsurprising that their point estimates for the incidence probabilities are overestimated and that their CPs in Scenario 2 are much lower than those in Scenario 1. In contrast, the performance of {MAGEC} in Scenario 2 is almost identical to its performance in Scenario 1 for both interval and point estimates, demonstrating its robustness of censored data.

In a more extreme scenario where 40\% of data are censored (Scenario 3), the proposed method ({MAGEC}) performs well, with little information loss compared with Scenarios 1 and 2. However, all other estimators of AE incidence {lead} to inferior CP due to an increased percentage of censoring. The point estimations obtained from PEM, LRM, NAM, and RVE in Scenario 3 are more biased than those obtained from these methods in Scenario 2. Based on the MAD and RMSE, there {are} larger deviations from the true values of the incidence probabilities compared with those in Scenario 2. {With an increasing total number of studies}, their CPs in Scenario 3 are much lower that those in Scenario 1 and 2. {When the percentage of censoring reaches 60\%, the performance of {MAGEC} is still the best among all methods.} { When we incorporate right-censored AE data, MAGEC consistently provides the lowest errors across all settings. In Scenario 6 with censoring possibly on both sides, the results from four other methods are unreliable.}
{For example, the observed errors sometimes are smaller in Scenario 6 (e.g., MAD = 0.015 in PEM) than in Scenario 3 or 5 for one-sided censored data alone (e.g., MAD = 0.033 or 0.020 in PEM), when combining two wrongs makes a right because the analysis of neglecting the left-censored or right-censored data tends to be biased in the opposite direction.}
However, as illustrated in Table \ref{tab:2}, the smaller error cannot be warranted across different sample size settings. Overall, the {MAGEC model} yields not only more stable and superior coverage, but also unbiased estimates of the incidence probabilities in all {six} scenarios {especially with censored event data}.
{The proposed method, handling both observed and censored data took, around 20 seconds in the absence of censoring and 50-70 seconds with an increasing percentage of censoring. In contrast, the other four methods, which only consider observe data, completed the task in less than one second. }

Across all scenarios considered above, {MAGEC} outperforms the other four methods in estimating incidence probabilities when AEs have low incidence and when a high proportion of AEs are censored. Furthermore, the quality of an estimator can be measured by its efficiency, which is defined as the asymptotic variance of an estimator \citep{casella2002statistical}. The larger the variance, the lower the efficiency of an estimator. Here, the asymptotic relative efficiency (RE), defined by the reciprocal of the ratio of the asymptotic variances of two unbiased estimators, is given to examine the amount of information loss from informative censoring. There {is}  a 10\%-20\% inflation in the RMSE from 0\% to 40\% censored data, suggesting limited efficiency loss of {MAGEC} in incidence estimation due to censoring. In summary, the proposed method consistently achieves a reasonable performance in estimating the incidence rates at different probability levels of AE incidence in the presence of censored data.

\section{Application} % section 4
\label{sec4}

In this section, we apply the proposed Bayesian method to modeling a meta-analysis of Grade 3 or higher AEs with censored information \citep{wang2019treatment}. The goal is to evaluate the incidence probabilities of pneumonitis (referring to inflammation of lung tissue) in two PD-1 and three PD-L1 inhibitors in a meta-analysis of 125 clinical studies. Pneumonitis is a typical serious AE not only leading to hospital admission and potentially permanent lung damage or death, but also resulting in treatment discontinuation \citep{chuzi2017clinical, zhong2020immune}. Such kind of inflammatory or immune-related AEs are of special interest for cancer immunotherapy. Specifically, around 35\% of grade 3 or higher AEs of pneumonitis across studies were incompletely reported.

We compare six models to determine the best model for grade 3 or higher pneumonitis data. The linear combination in (\ref{eq3}) could entail one or more of the following terms: the between-study random effects ($\alpha$), drug or drug-dose random effects ($\eta$), and cancer type random effects ($\zeta$). Model 1 (M1) is the simplest model with only study effects. Model 2-4 (M2-M4) consider adding cancer effect, drug effect, and drug-dose effect to the M1, respectively. With both cancer effect and drug effect, Model 6 (M6) is more complex than Model 5 (M5) because it also takes dose level into consideration. % Model intro
To specify the priors for each random effect in the proposed logistic model, we assume normal distributions
\begin{align}
    \alpha \sim N(0, \sigma^2_{\alpha}), \quad
    \eta \sim N(0, \sigma^2_{d}) \text{ or }  \sim N(0, \sigma_{ad}^2),  \quad 
    \zeta \sim N(0, \sigma^2_{c}), \nonumber
\end{align}
with half-Cauchy prior distributions for the standard deviation parameters and the scale parameter $A=25$ \citep{gelman2008weakly}.

 \begin{table}[htb]
 \begin{center}
     \scalebox{0.85}{\begin{tabular}{ccccccc} 
         \hline
         \hline
        Model & $\sigma_\alpha$ & $\sigma_c$ & $\sigma_{d}$ & $\sigma_{ad}$ & DIC & WAIC \\ 
        \hline
        M1 & 0.714 [0.415, 1.100] & - & - & - & 243.7  & 223.2 \\ % study 
        M2 & 0.494 [0.060, 0.927] & 0.697 [0.186, 1.864] & - & - & 248.5  & 227.2 \\ % study, cancer 
        M3 & 0.701 [0.401, 1.080] & - & 0.357 [0.021, 1.650] & - & 241.1 & 222.0 \\ % study, drug 
        M4 & 0.704 [0.396, 1.084] & - & - & 0.223 [0.010, 0.822]   & 244.0 & 223.9 \\ % study, drug-dose 
        M5 & 0.425 [0.037, 0.865] & 0.739 [0.240, 1.977] & 0.458 [0.038, 1.904] & - & \textbf{237.0} & \textbf{217.1} \\ % study, cancer, drug
        M6 & 0.410 [0.032, 0.876] & 0.739 [0.232, 1.959] & - & 0.321 [0.020, 0.899] & 238.7 & 218.2\\ % study, cancer, drug-dose
        \hline
     \end{tabular}}
     \caption[]{Model Comparison: heterogeneity between study ($\sigma_\alpha$), cancer type ($\sigma_c$), and drug ($\sigma_d$) or drug-dose ($\sigma_{ad}$),  deviance information criterion (DIC), and widely applicable information criterion (WAIC) from modeling observed and censored grade 3 or higher AE (pneumonitis) data.} 
     \label{tab:3}
 \end{center}
 \end{table}

\begin{figure}[!htb]%[htbp]
	\centering
	\centering \vspace{-25pt}
	\includegraphics[scale=0.6, trim={0cm .8cm 0cm 1.8cm},clip]{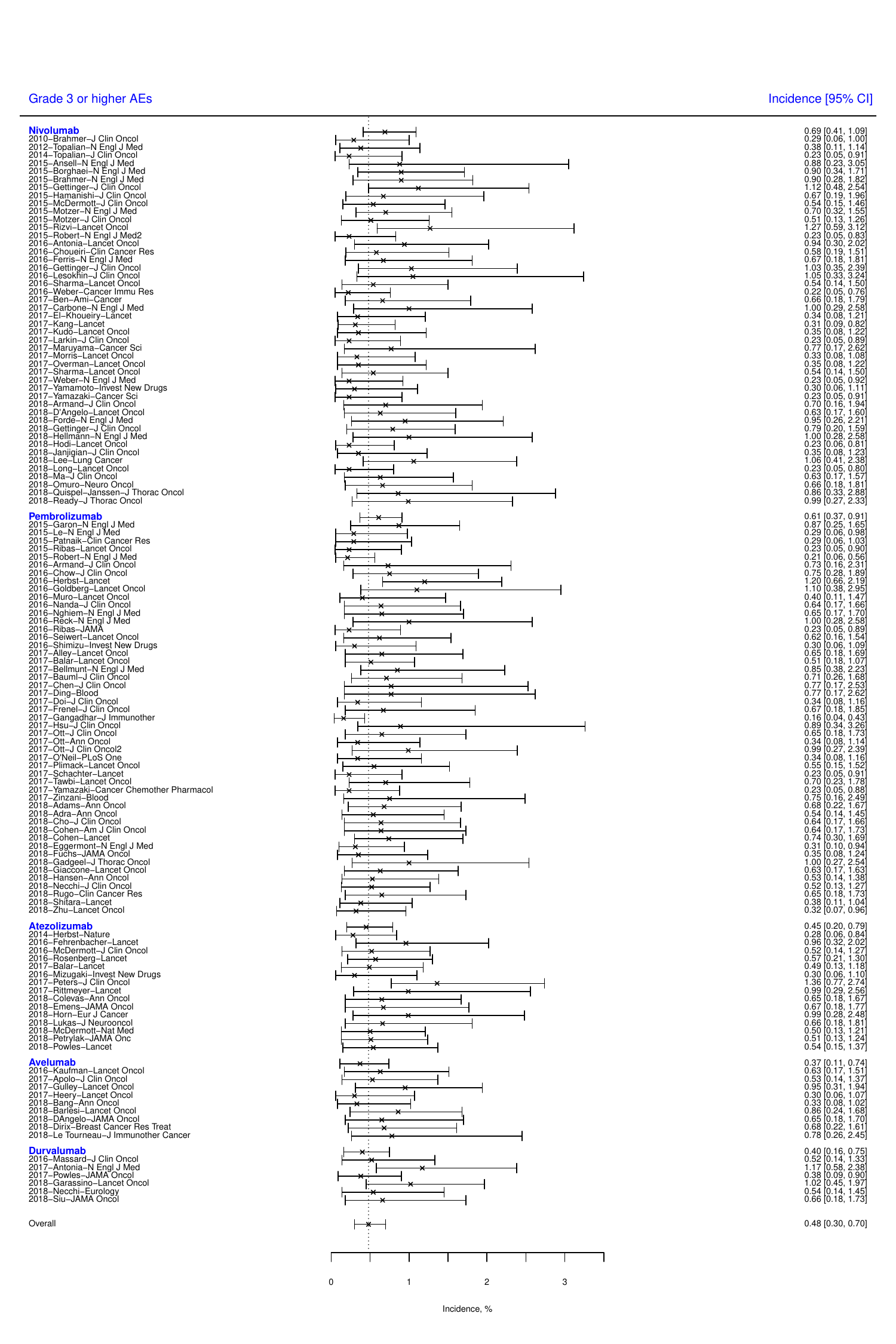}
	\caption{Incidence of grade 3 or higher AE (Pneumonitis) by study}
	\label{fig:4.2}
\end{figure} 

All six models are implemented in the statistical software R and JAGS \citep{plummer2003jags} using MCMC algorithms to generate samples from the posterior distribution of the parameters of interest. We ran three parallel chains for the model. For each MCMC chain, after discarding the burn-in period of 30,000 iterations, the 3 chains showed good mixing and successful convergence to the target distribution. We eventually obtained 10,000 posterior samples per chain by retaining one sample out of three. Based upon the traceplot for each parameter of interest, all chains are close to each other and the target distribution. Furthermore, in terms of the potential scaler reduction statistic, also known as \texttt{Rhat}, the values of all key parameters are at 1, suggesting convergence. The 30,000 posterior samples of model parameters such as incidence probabilities of cancer effects and drug effects  were saved for inference. 

The model assessment results are summarized in Table \ref{tab:3}. Model 5 containing study, cancer and drug effects was selected based upon the smallest DIC and WAIC among all models considered. According to the subgroup analysis of incidence probability of AE by drug, there were no significant differences in the incidence among different dosing schedules for PD-1/PD-L1 drugs. {Figure \ref{fig:4.2} displays the estimated incidence probabilities, along with their 95\% credible intervals, for grade 3 or higher pneumonitis across all 125 clinical trials. The vertical dashed line {in Figure \ref{fig:4.2}} is the overall incidence probability of grade 3 or higher pneumonitis (0.48\%; 95\% CI, 0.30\%-0.70\%) across all studies.  By contrast, if the censored outcomes were treated as missing at random, the estimated incidence rate would be biased and overestimated by $12.5\%$.{If we further ignored the covariate information and analyzed the case study using four other frequentist methods in simulation study, the overall incidence probability estimates would increase more than 60\%. } The case study results are consistent with our simulation studies, in that existing methods which rely only on reported counts (which by definition were not subject to censoring) tend to overestimate the incidence rates. Our proposed method avoids this upward bias, suggesting a more favorable safety profile for the PD-1/PD-L1 drugs. These findings demonstrate that our proposed method can play a critical role in providing a more accurate understanding of the risk/benefit ratio for new therapies with serious but low-frequency adverse events.

\section{Discussion}
\label{sec5}

In this paper, we proposed a simple Bayesian hierarchical modeling framework in the meta-analysis setting when the study-level event rates could be incompletely reported. Our Bayesian approach, named MAGEC, is efficient to handle censored safety data without requiring the intensive computation that multiple imputation methods rely on for full Bayesian modeling.
Further, we demonstrated the superior performance of our proposed method in simulations. Specifically, simulation results showed that the proposed Bayesian approach generated unbiased estimations of drug effects with limited information loss. Moreover, the proposed method was robust to rare events, which are commonly encountered in data analysis of AEs. 
Note that the definition of rare events here is in a statistical sense following the previous works in the meta-analysis \citep{bhaumik2012meta, liu2014exact}, rather than in the pharmacovigilance setting \citep{council1999guidelines}.
Finally, we illustrated the implementation with a real data application, while employing the proposed method can effectively reduce the over-estimation bias towards the incidence rate of pneumonitis in cancer patients with the treatment of immunotherapy. 

Special consideration must be given to incomplete reporting of AEs to avoid incorrect statistical inference. In general, there are mainly three concerns when conducting safety meta-analyses: (1) AE incidences are heterogeneous across studies, (2) AEs of interest could be rare, and (3) AEs are  incompletely or inconsistently reported  \citep{berlin2013meta,stoto2015drug}. While various researchers have proposed different methods to address the first two concerns, the third concern is largely overlooked in methodology development.
Naively applying standard methods in the presence of missing or censored data can result in bias or incorrect conclusions \citep{little2019statistical}.  
In our motivating application, clinical trial reporting standards only require that AEs be included in the final trial report if they exceed a pre-specified cut-off that is specific to that study; this means that rare AEs that occur less frequently than the threshold are omitted and can be considered as left-censored data. For some older clinical trials, the trial cutoff could be (implicitly) as large as the sample size, while only efficacy outcomes were reported in the publication with no mention of AEs \citep{golder2016reporting}. Particularly for newly developed drugs, we believe that such completely censored reporting or related publication bias should not be a major concern. Since April 18, 2017, it is mandated that all studies registered with \url{ClinicalTrials.gov} must report AE counts that exceed a pre-specified cutoff (e.g. $5\%$) for each study arm within one year after the primary completion date. Nevertheless, the problem of incomplete AE reporting with study-specific cutoff, as this paper attempts to address, is still unavoidable for meta-analysis of drug safety with this recent reporting requirement. 

The proposed Bayesian MAGEC modeling framework estimates the incidence probabilities using a one-stage approach. In contrast, a meta-analysis of binary data is usually conducted using a two-stage approach in the literature \citep{deeks2002issues, simmonds2005meta} - the summary statistics are first separately calculated for each trial and then combined by an appropriate meta‐analysis model. However, this two-stage approach is likely to perform poorly in the first stage of each study due to the rarity of events \citep{burke2017meta}. Alternatively, a one-stage approach is preferred as it delivers  more exact statistical inference \citep{debray2013individual}. 

Although incomplete reporting of AE data and its impact on biasing meta-analysis results has been noted by several researchers, no statistical approach has yet been proposed to systematically address this issue. Our Bayesian MAGEC model represents the first such attempt, and we present it in a form of simplicity to illustrate the concept. {One limitation of our proposed method is that MAGEC requires a reasonable number of eligible studies to be included in meta-analysis of drug safety. In cases with very few studies, it becomes important to apply proper adjustments to enable stable estimation for random effects \citep{mathes2018comparison,zhou2022statistical}. In the Bayesian modeling setting, a robust prior for the standard deviation parameter, such as the half-Cauchy prior with heavy tails, is preferred for the between-study random effect \citep{gelman2006prior}. Another limitation of MAGEC is the assumption that the censored safety data exhibit a non-random missingness pattern. This assumption is analogous to the standard censoring types often encountered in survival data analysis, where we assume that the data are censored in a non-informative manner.} In the real-data application, we only demonstrated the case of left-censored grade 3 or higher AEs when they happened at a frequency lower than pre-specified cutoff values. However, the proposed framework can be extended to joint modeling of AEs of different severity grades also with right-censoring using multinomial distribution. {Additionally, as many AE types are clustered in nature, multivariate modeling, rather than univariate modeling, could address the AE correlation and explore high-risk subgroups.} 
Furthermore, it's worth noting that our proposed method can be applied to the meta-analysis of high-dimensional genomic data \citep{hill2020genetic}, in which a large number of genes are evaluated to estimate the mutation rate in the panels across studies.  For such an extension, information on some mutations could be also censored due to low frequency, which should be considered in the model using a pre-specified cutoff value determined by gene selection criteria.  
{From the modeling perspective, the high-dimensional outcome can be modeled directly using a multivariate normal distribution with a covariance matrix describing the genome-wide association. In implementation can be simplified with pre-existing genomic knowledge or specific covariance structure such as compound symmetry or first-order autoregressive structure. Additional parameters in the covariance matrix can be sampled from their conditional distributions using MH algorithm. Compared to the current simulation setting, computation for high-dimensional censored data demands sufficient resources in computational power and could be more time-consuming, but we also expect a better performance due to information borrowing in the Bayesian framework.} Lastly, our model implementation strategy could also be applied to analyze other censored data structures, including time-to-event data, count data, and ranking data, in addition to binomial data \citep{johnson2013bayesian}. By providing a statistical framework to systematically address incomplete AE reporting, we hope that the Bayesian MAGEC model will facilitate more accurate and reliable meta-analyses in the future.

\section*{Data Availability Statement}
The data and code to reproduce real-data analysis are available from \url{https://github.com/xinyue-qi/Meta-analysis-of-Censored-AEs}.

\section*{Acknowledgement}
This work is supported in part by NIH CCSG Grant P30CA016672. 

\setcounter{figure}{0}
\setcounter{table}{0}
\renewcommand{\thefigure}{C.\arabic{figure}}
\renewcommand{\thetable}{B.\arabic{table}}

\newpage
\section*{Appendix: \texttt{JAGS} model specification}
Here we provide the \texttt{JAGS} model specification for the application in Section \ref{sec4}.
\begin{small}
\begin{Verbatim}[tabsize=6]
# application: study, cancer, drug effect included
model{
    for (j in 1:J1){
        Y[j] ~ dbin(theta[j], N[j])
        logit(theta[j]) <- theta.s[study[j]] + theta.d[drug[j]] + theta.c[cancer[j]]
    }
    for (j in 1:J2){
        Z[j] ~ dbern(p[j])
        p[j] <- pbin(cut[j], theta[j+J1], N[j+J1])  #Y<=cut
        logit(theta[j+J1]) <- theta.s[study[j+J1]] + theta.d[drug[j+J1]] + theta.c[cancer[j+J1]]
    }
    for (i1 in 1:n.study){
        theta.s[i1] <- mu.study + sigma.study*sn.study[i1]
        sn.study[i1] ~ dnorm(0,1)
    }
    mu.study ~ dnorm(0, .0001)
    sigma.study ~ dt(0, a, 1)T(0,) # a=1/A^2
    for (i2 in 1:n.drug){
        theta.d[i2] <- mu + sigma.drug*sn.drug[i2] 
        sn.drug[i2] ~ dnorm(0,1) 
    }
    mu ~ dnorm(0, .0001)	
    sigma.drug ~ dt(0, a, 1)T(0,) 
    for (i5 in 1:n.cancer){
        theta.c[i5] <- mu.cancer+sigma.cancer*sn.cancer[i5]
        sn.cancer[i5] ~ dnorm(0, 1)
    }
    mu.cancer ~ dnorm(0, .0001)
    sigma.cancer ~ dt(0, a, 1)T(0,)
}	
\end{Verbatim}
\end{small} %footnotesize

\newpage
\bibliographystyle{unsrtnat}
\bibliography{arXiv_v3}

\end{document}